\documentclass[aps,prl,floatfix,amsmath,amssymb,twocolumn,superscriptaddress]{revtex4}
\usepackage{latexsym}
\usepackage{amsfonts}
\usepackage{amssymb}
\usepackage{amsbsy}
\usepackage{newtxtext,amsmath}
\usepackage{dsfont}
\usepackage{graphicx}
\usepackage{float}
\usepackage{bm}
\usepackage{natbib}

\renewcommand{\figurename}{Fig.}

\renewcommand{\refname}{Ref.}

\newcommand{\expressionname}{Eq.}
\newcommand{\expressionsname}{Eqs.}

\begin{document}

\title{Classical Liquids in Fractal Dimension}

\author{Marco Heinen}
\email[]{mheinen@caltech.edu}
\affiliation{Division of Chemistry and Chemical Engineering, California Institute of Technology, Pasadena, California 91125, USA}

\author{Simon K. Schnyder}
\affiliation{Department of Chemical Engineering, Kyoto University, Kyoto 615-8510, Japan}

\author{John F. Brady}
\affiliation{Division of Chemistry and Chemical Engineering, California Institute of Technology, Pasadena, California 91125, USA}

\author{Hartmut L\"{o}wen}
\affiliation{Institut f\"{u}r Theoretische Physik II, Weiche Materie, Heinrich-Heine-Universit\"{a}t D\"{u}sseldorf, 40225 D\"{u}sseldorf, Germany}

\date{\today}

\begin{abstract}
We introduce fractal liquids by generalizing classical liquids of integer dimensions $d=1,2,3$ to a non-integer dimension $d_l$.
The particles composing the liquid are fractal objects and their configuration space is also fractal, with the same dimension.
Realizations of our generic model system include microphase separated binary liquids in porous media, and
highly branched liquid droplets confined to a fractal polymer backbone in a gel.
Here we study the thermodynamics and pair correlations of fractal liquids by computer simulation and semi-analytical statistical mechanics.
Our results are based on a model where fractal hard spheres move on a near-critical percolating lattice cluster.
The predictions of the fractal Percus-Yevick liquid integral equation compare well with our simulation results.
\end{abstract}

%\pacs{61.20.Gy %Theory and models of liquid structure,
%      61.20.Ja %Computer simulation of liquid structure, 
%      61.43.Hv %Fractals; macroscopic aggregates 
%      }
\maketitle

The liquid state, an intermediate between gas and solid, exhibits short-ranged particle
pair correlations in isotropic shells around a tagged particle \cite{Hansen_McDonald2006}.
While the shell structure is lost for gases, solids exhibit long-ranged correlations and anisotropy.
Particle correlations are accessible by experiments \cite{Kegel2000,Royall2008} and contain valuable
information about the particle interactions.
A fundamental task in theory and computer simulation of the liquid state is to predict particle
correlations for given interactions.
In this respect, one particularly successful approach is liquid integral equation
theory \cite{Hansen_McDonald2006, Caccamo1996}.

Molecular and colloidal liquids can be restricted to one or two spatial dimensions
\cite{Alba-Simionesco2006, Schoen2007} by confining them on substrates
\cite{Deutschlaender2013}, at interfaces \cite{Zahn2000}, between plates \cite{Neser1997, Steward2014},
in channeled matrices \cite{Hahn1996} or optical landscapes \cite{Wei2000}.
The thermodynamic properties change accordingly \cite{Franosch2012}
and the systems are called one-dimensional (1D) or two-dimensional (2D) liquids.
While these classical examples of confinement involve integer-dimensional
configuration spaces, there are also cases where liquids are confined in
porous media \cite{Kurzidim2009, Kim2011a, Skinner2013, torquato2002random},
or along quenched polymer coils \cite{Barrat_Hansen2003}, and where the configuration
space exhibits non-integer (fractal) dimension at suitable length scales.
Much work has been devoted to understanding the limiting cases of low and high density,
namely the motion of a single particle on a fractal \cite{Metzler2000, benAvraham_Havlin2000, Seeger2009, Muelken2011, Hofling2013}
and the structure of a fractal aggregate itself, corresponding to an arrested high-density particulate
system \cite{Wong1992, Meakin1999, Sorensen2001, Zhao2005, Poon1995}.
In all these situations, particles interact in the embedding integer-dimensional space
and interaction energies depend on Euclidean particle distance.

\begin{figure}
 \fbox{\includegraphics[width=.85\columnwidth]{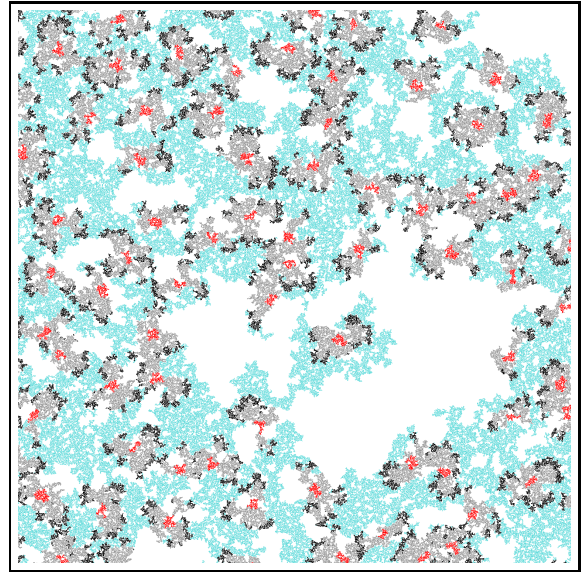}}
 \caption{(Color online) Snapshot of a Monte Carlo simulation where
 $300$ fractal particles are located on a percolating cluster shown in blue.
 The dimension of each particle and of the cluster is $d_l = 1.67659$,
 and the chemical distance diameter of each particle is $\sigma = 300a$ (\textit{c.f.}, \figurename{}~\ref{fig:Metric}).
 Red, black and gray pixels are occupied by particles.
 Red: Particle center regions (chemical distance to a particle center is less than $0.1\sigma$).
 Black: Particle rim regions (chemical distance to a particle center is more than $0.4\sigma$ and less than $0.5\sigma$).
 Gray: Regions between particle centers and rims.
 Every pixel in the figure corresponds to one vertex in the simulation.
 One quarter of the whole simulation box is shown, containing here $72$ particle centers.}
 \label{fig:snapshot}
\vspace{-.75em}
\end{figure}

In this letter, we break new ground by considering \emph{fractal particles} in a fractal configuration space,
both of the same non-integer dimension. An alternative model in which the particle dimension differs from
the configuration-space dimension is briefly mentioned near the end of the letter.
Figure ~\ref{fig:snapshot} is a snapshot from one of our Monte Carlo (MC) simulations. 
The interaction between two particles is not described by a pair potential in Euclidean space,
but instead inherits fractal character from the particles and their configuration space.
Unifying the two complementary fields of classical liquid state theory and fractal structures,
we refer to the explored, dense disordered phases as \emph{'fractal liquids'}.
As opposed to continuum-mechanical descriptions of fluid flow on fractals \cite{Onda1996,Moreles2013}
or liquids forming fractal stream networks \cite{Tarboton1988,Tomassone1996},
here we calculate the correlations of individual fractal particulates that constitute a liquid. 
We develop basic concepts for statistical mechanics of fractal liquids by
presenting MC simulations along with an accurate semi-analytical fractal liquid integral equation approach.
Our model system is a fractal hard sphere liquid, the analogue of the simplest generic model liquid in integer dimensions.
In the present proof of principle, we restrict our studies to the liquid state.
Determination of the full phase diagram of fractal particles in fractal space is a task that should be
tackled in future work. In particular, it will be interesting to study whether or not symmetry breaking and
crystallization occur in fractal spaces. 

Fractal liquids are not only interesting from a theoretical perspective, but are also found in nature.
Porous media filled with phase-separated binary liquids such as water and oil appear
in natural oil or gas reservoirs, and are produced during hydraulic fracturing in oil drilling.
If the oil droplets exceed the porosity length scale in size, they adapt to the fractal shape of the void space.
Such droplets constitute fractal 'particles' in a fractal configuration space set by the porous medium.
While the individual water and oil molecules are confined to the void space, it
is important to note that the mesoscopic water or oil droplets are not necessarily confined:
If the size of the void space greatly exceeds the droplet size, then the droplets move 
in a (quasi) unbounded configuration space of their own non-integer dimension.
Here we study the collective bulk-phase thermodynamics of such droplets, which are the fractal liquid constituents.  
Other possible model systems include different mesoscopic 'particles',
such as floppy slime molds \cite{Baumgarten2014} or droplets confined to fractal polymer aggregates in a gel.

To model a fractal liquid we first need a fractal-dimensional configuration space with a suitable distance measure. 
A viable choice is a large set of vertices on a discrete lattice for a fractal lattice liquid simulation.
From a regular square lattice with a number $W$ of vertices and periodic boundary
conditions in embedding 2D Euclidean space, we randomly remove vertices until $V$ vertices are left.
According to percolation theory
\cite{Stauffer_Aharony1994, benAvraham_Havlin2000}, in the limit $W\to\infty$ there is a critical percolation threshold $p_c = V/W$
at which a fractal-dimensional percolating cluster extends to infinite length scales.
For the square lattice, a threshold of $p_c = 0.5927\ldots$ has been reported \cite{Newman2000}.

\begin{figure}
\includegraphics[width=.45\columnwidth]{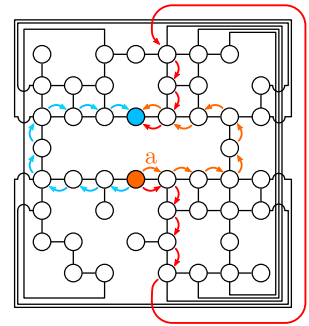}
 \caption{(Color online) Lattice schematic.
 From an $8 \times 8$ square lattice with periodic boundaries, $17$ vertices have been removed.
 Circles are the remaining vertices and black lines are the bonds of length $a$.
 The chemical distance between the orange and the blue filled circle is $l = 8a$
 along any of the three paths indicated by red, blue and orange arrows.
 The corresponding Euclidean distance is $r = 2a$.}
 \label{fig:Metric}
\vspace{-.75em}
\end{figure}

Throughout this work, all distances $l$ between vertices are evaluated as shortest distances on the percolating cluster,
using our own implementation of what is essentially equal to \mbox{Dijkstra's} algorithm \cite{Dijkstra1959}.
The distance between two vertices is $l = a$ if and only if these vertices are connected by a bond in one of the two Cartesian
directions of the embedding space.
Illustrated in \figurename{}~\ref{fig:Metric}, this distance measure is commonly referred to as
'chemical distance' and the corresponding non Euclidean metric is known as the 'taxicab metric'.

\begin{figure}
\includegraphics[width=.56\columnwidth,angle=-90]{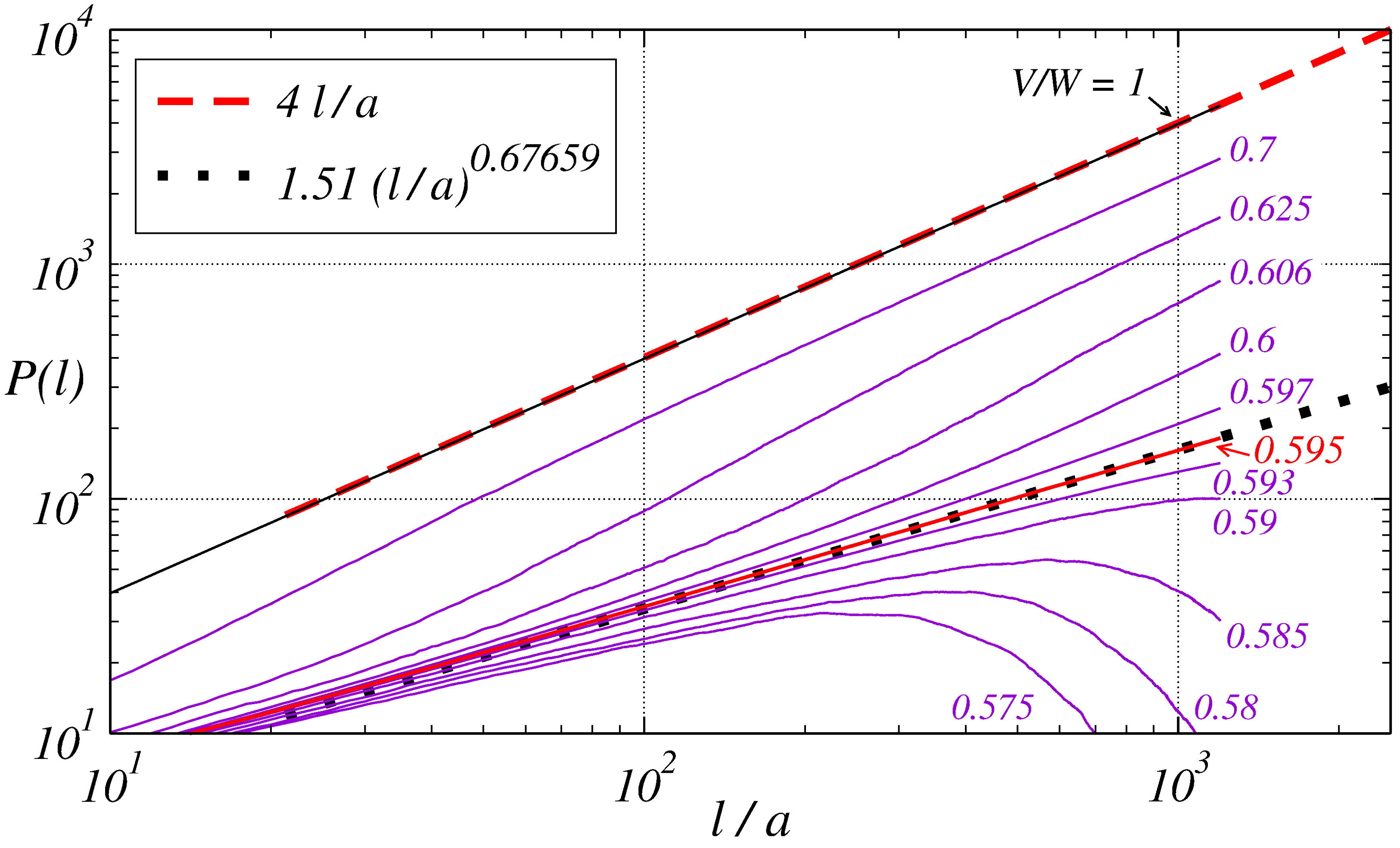}
\vspace{-.5em}
 \caption{(Color online) Fractal clustersize scaling.
 Solid curves indicate average numbers, $P(l)$,
 of vertices with chemical distance $l$ from a tagged vertex,
 in the largest cluster remaining from a lattice of initially $W = 3000 \times 3000$ vertices.
 Results are for various numbers $W-V$ of removed vertices, with ratios $V/W$ as indicated.
 Upper dashed line: $P(l) = 4 ~ l/a$, corresponding to a 2D lattice.
 Lower dotted line: $P(l) = 1.51 ~ {(l/a)}^{0.67659}$, corresponding to a $1.67659$-dimensional lattice.
 Our simulations are for $V/W = 0.595$. }
 \label{fig:Exponent}
\vspace{-.75em}
\end{figure}

The lattice in \figurename{}~\ref{fig:Metric} features the essential
properties of the lattices in our simulations, except of their much larger size of $W = 3000 \times 3000$, $V = 5~355~000$.
The ratio $V/W = 0.595$ is $0.4\%$ larger than $p_c$, which
ensures the presence of a percolating cluster in the finite system.
We retain only those vertices that belong to the percolating cluster.
All other vertices, either in smaller clusters or isolated, are deleted.
An average count of $3 \times 10^6$ connected vertices remains.
As shown in \figurename{}~\ref{fig:Exponent},
the average number $P(l)$ of vertices with chemical distance $l$ from a tagged vertex grows as a non-integer power of $l$.

The so-called 'spreading dimension' $d_l = 1.67659\ldots$ of the percolating cluster
is the relevant dimension here. It has been determined by MC simulations \cite{Zhou2012} which
agree with our result $P(l) \propto l^{d_l-1}$.
Note that $d_l$ differs from the more commonly reported fractal cluster dimension $d_f = 91/48$
\cite{benAvraham_Havlin2000, Stauffer_Aharony1994, Zhou2012},
because chemical distance scales as a non-integer power of Euclidean distance. 

We define a fractal particle as the set of vertices that are within a distance $l \leq \sigma/2$ from a center vertex.
There is exactly one center vertex per particle and we refer to $\sigma$ as the particle diameter.
Every particle occupies at least one vertex pair with distance $l = \sigma$ at all times.
Particles obey a no-overlap constraint, prohibiting configurations in which any two particle centers have a distance of $l \leq \sigma$.
We choose $\sigma = 300a$, with the consequence that the particles themselves are fractal objects
(\textit{c.f.} \figurename{}~\ref{fig:Exponent}).
As the snapshot in \figurename{}~\ref{fig:snapshot} shows, the particles are highly anisotropic in the embedding 2D space.
However, in fractal space and taxicab metric, the particles and their no-overlap interactions are perfectly isotropic.
In spite of their polymorphy in embedding space, all particles are \emph{indistinguishable} for the purpose of our simulation.
Hence the particles are fractal-dimensional analogues of monodisperse hard spheres,
and one may refer to the simulated system as the \mbox{$1.67659$-dimensional} hard sphere liquid.

We probe the thermodynamic equilibrium state by MC simulation.
Particles are moved one at a time, and every move proceeds as follows:
After deleting a random particle, we pick a random vertex $v_1$ globally from the cluster, which is a candidate to become
a vertex at the rim of the displaced particle.
We then pick a random vertex $v_2$ at distance $l = \sigma$ from $v_1$, and 
a random vertex $v_3$ at distance $l = \sigma/2$ from both $v_1$ and $v_2$.
Vertex $v_3$ is the candidate to become the displaced particle's center vertex.
If there is a center vertex of another particle at distance $l \leq \sigma$ from $v_3$,
then the move is rejected and the original particle restored.
Otherwise, the move is accepted.

The simulation starts with a random configuration of potentially overlapping particles with diameter $\sigma = 10a$.
In the initial simulation stage $\sigma$ is gradually inflated to $300a$, which facilitates finding a dense configuration without overlaps.
After the initial inflation and equilibration the production stage is entered,
and the average number $N(l)$ of particles with center-to-center distance $l$ is recorded.
Binning the function $N(l)$ with a bin width of $5a$ reduces scatter in the data.
Inside a fractal control volume, the fractal packing fraction $\phi$ is measured as the average number ratio of
vertices occupied by particles to the total number of contained vertices.
The control volume is a set of vertices with distance $l \leq 1500a$ from a center vertex.
It is randomly moved during the simulation and contains at least one vertex pair with distance $l = 3000a$ at all times. 
We compute the chemical distance distribution function 
\begin{equation}\label{eq:gr_simul}
g(l) = c~\times~N(l)~\times~{(a/l)}^{d_l - 1},
\end{equation}
with the $l$-independent factor $c$ chosen such that $g(l) \to 1$ for large $l$.
Function $g(l)$ is the analogue of the radial distribution function $g(r)$
of isotropic liquids in integer dimensions \cite{Hansen_McDonald2006}.
Our simulations require long runtimes due to the numerically expensive distances calculations.
The fractal cluster's ramified structure severely complicates any performance improvement,
except for parallel execution of statistically independent runs and subsequent averaging.

\begin{figure}

\includegraphics[width=.45\columnwidth,angle=-90]{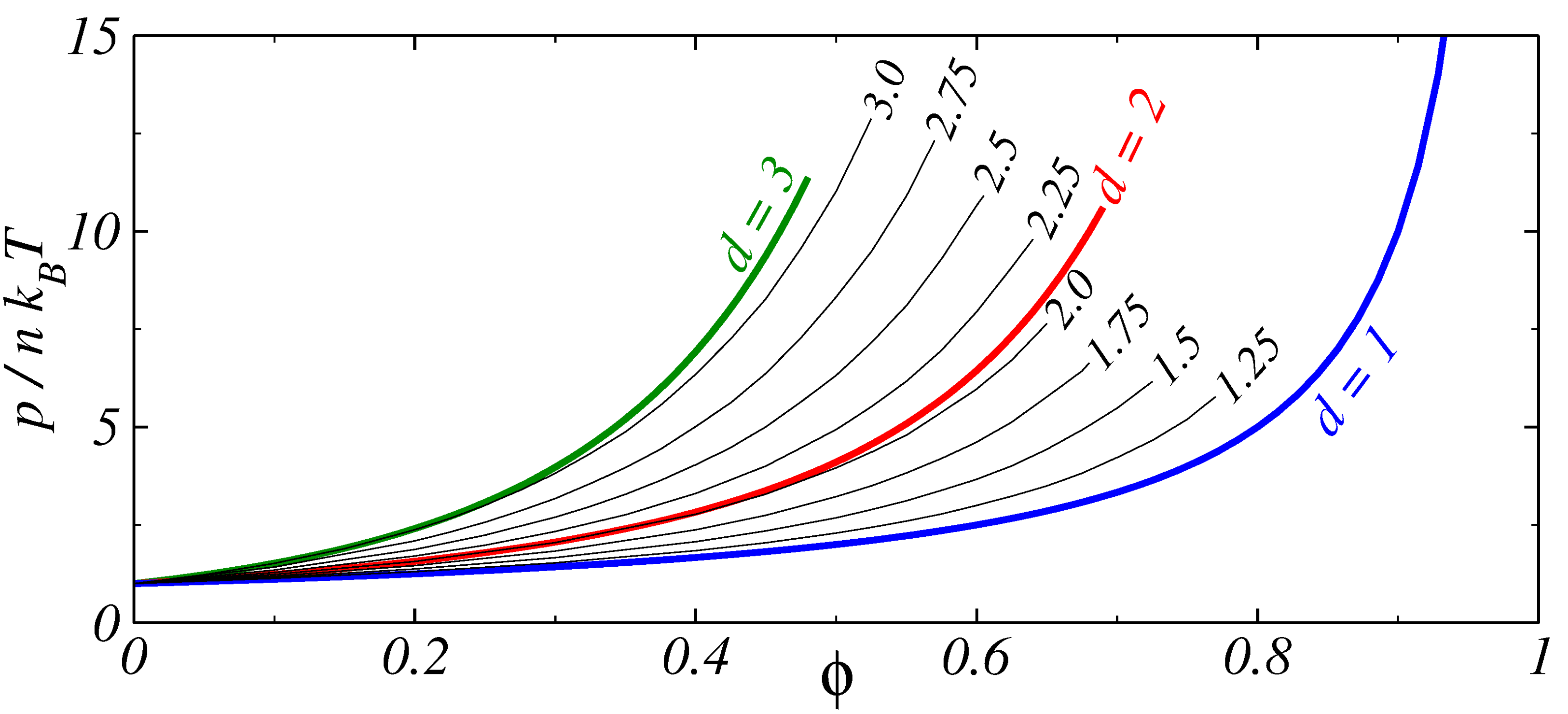}
\vspace{-.75em}
 \caption{(Color online) Equations of state for hard spheres in dimensions from $d = 1$ to $d = 3$.
 Thick blue curve: Exact (Tonks gas / PY) result for $d = 1$.
 Thick red curve: Kolafa-Rottner \cite{Kolafa2006} virial series with 15 coefficients for $d=2$.
 Thick green curve: Carnahan-Starling equation of state for $d=3$.
 Thin black curves are the PY virial pressure results for dimensions as indicated.}
 \label{fig:EOS}
\vspace{-.75em}
\end{figure}

The Ornstein-Zernike equation for a homogeneous and isotropic liquid in integer-dimensional Euclidean space reads
\begin{equation}\label{eq:OZ}
g(r) = 1 + c(r) + n \int d^d \boldsymbol{r}' [g(r') - 1]~c(|\boldsymbol{r}-\boldsymbol{r}'|),
\end{equation}
where $n$ is the particle number density, $c(r)$ is the direct correlation function \cite{Hansen_McDonald2006},
$\boldsymbol{r}$ is a $d$-dimensional vector with Euclidean norm $r = |\boldsymbol{r}|$, and $d^d\boldsymbol{r}'$
is an infinitesimal volume element at position $\boldsymbol{r}'$.
In conjunction with the no overlap constraint $g(r\leq\sigma)=0$ and the approximation $c(r > \sigma)=0$, \expressionname~\eqref{eq:OZ}
constitutes the Percus-Yevick (PY) integral equation for $d$-dimensional hard spheres \cite{Percus1958},
which can be systematically derived by functional Taylor expansion \cite{Hansen_McDonald2006}.

We solve the PY equation by means of a spectral solver \cite{Heinen2014}.
Our numerically efficient algorithm for the convolution-type \expressionname~\eqref{eq:OZ}
is based on the Hankel transform pair
\begin{eqnarray}
\tilde{f}(q) &=& \frac{{(2\pi)}^{d/2}}{q^{d/2-1}} \int\limits_0^\infty dr~ r^{d/2} f(r) J_{d/2-1}(qr), \label{equ:Fourier_r_to_q}\\
{f}(r) &=& \frac{r^{1-d/2}}{{(2\pi)}^{d/2}} \int\limits_0^\infty dq~ q^{d/2} \tilde{f}(q) J_{d/2-1}(qr), \label{equ:Fourier_q_to_r}
\end{eqnarray}
for a $d$-dimensional isotropic function $f$, sampled on a logarithmic grid \cite{Talman1978, Rossky1980, Hamilton2000}.
In \expressionsname~\eqref{equ:Fourier_r_to_q} and \eqref{equ:Fourier_q_to_r}, $J_{d/2-1}(x)$
is the Bessel function of the first kind and order $d/2-1$.
Since $J_{d/2-1}(x)$ is analytic with respect to both $x$ and $d$,
the solution can be carried out formally also for non-integer dimensions.
Resulting pair-correlations and thermodynamic properties represent the analytic continuations of standard
PY theory with respect to the dimension.  

The capability of liquid integral equations to predict thermodynamic properties in arbitrary dimension
is exhibited in \figurename{}~\ref{fig:EOS}, featuring equations of state for hard spheres in various dimensions from $d=1$ to $d=3$.
Thin black curves in \figurename{}~\ref{fig:EOS} represent the $d$-dimensional hard sphere reduced virial pressure
%
%\begin{equation}\label{eq:ivir_press}
$p / n k_B T = 1 + 2^{(d-1)} \phi g(r = \sigma^+)$
%\end{equation}
%
with Boltzmann constant $k_B$, absolute temperature $T$, packing fraction $\phi = 2 \pi^{d/2} (\sigma/2)^d n / d \Gamma(d/2)$,
and $g(r)$ calculated in the PY scheme.   
Thick curves in \figurename{}~\ref{fig:EOS} are exact or nearly exact reference solutions for $d = 1, 2$ and $3$.
For $d = 1$, the PY result reduces to the exact (Tonks gas) solution $p / n k_B T = 1/ (1-\phi)$.
The good agreement of the PY-scheme with the numerically accurate result from \refname{}~\cite{Kolafa2006} for \mbox{$d=2$}
verifies the fidelity of the PY scheme for $1 \leq d \leq 2$.
Comparison to the Carnahan-Starling equation of state $p / n k_B T \approx (1 + \phi + \phi^2 - \phi^3) / (1-\phi)^3$ for \mbox{$d=3$}
reveals a decreasing PY-scheme accuracy for higher dimensions at large values of $\phi$.

\begin{figure}
\includegraphics[width=.82\columnwidth,angle=-90]{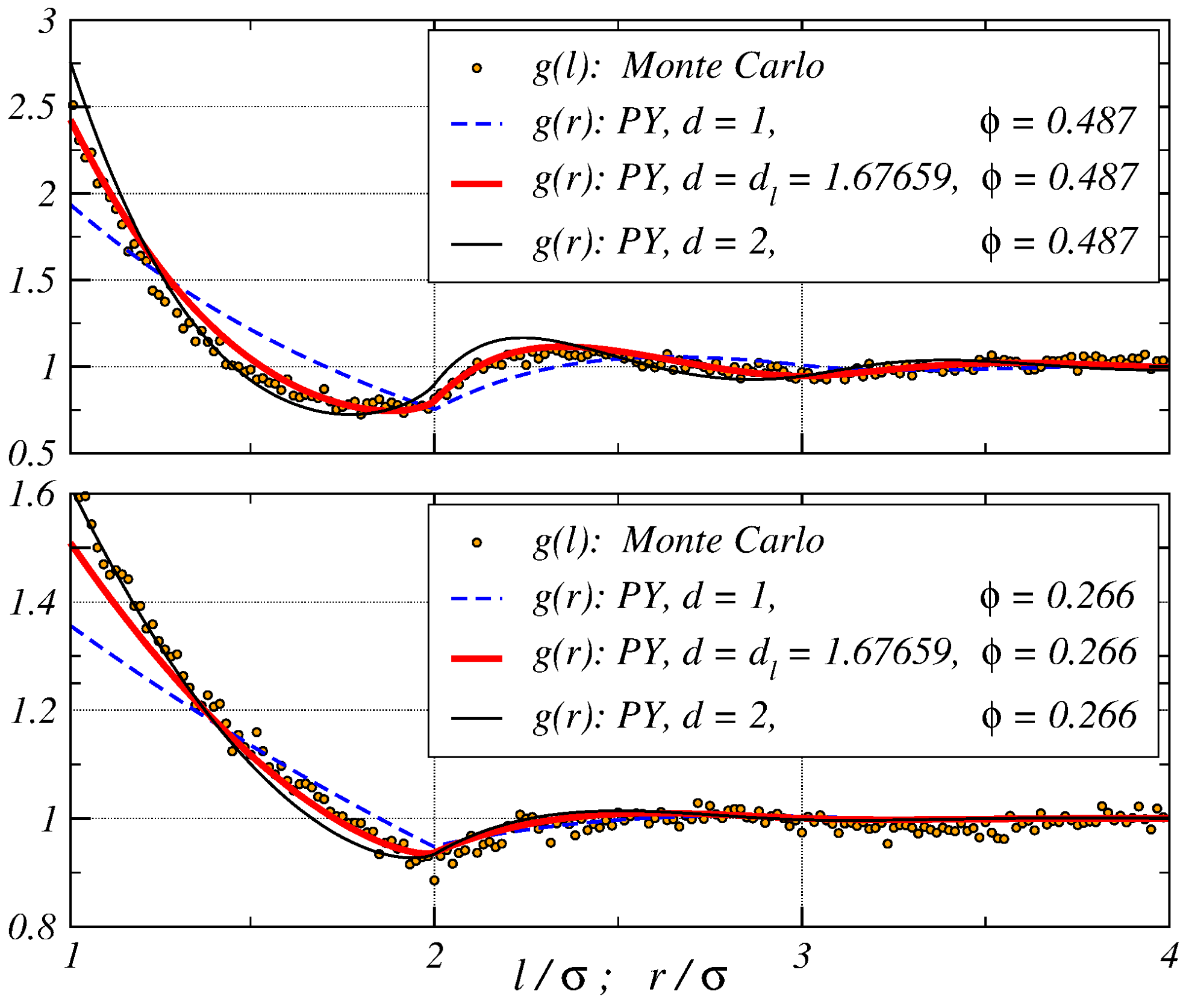}
\vspace{-.75em}
\caption{(Color online) Symbols: Chemical distance distribution functions $g(l)$ from our MC simulations.
Curves: PY solutions $g(r)$, for dimensions and (fractal) packing fractions as indicated.
Top and bottom panels are for $300$ and $150$ simulated particles, respectively.}
\label{fig:RDFs}
\vspace{-.75em}
\end{figure}

Figure~\ref{fig:RDFs} shows the functions $g(l)$ extracted from our simulations with $300$ and $150$ particles.
Also shown are the PY-scheme solutions $g(r)$ for
$d = 1$, $d = d_l = 1.67659$ and $d = 2$ and for $\phi = 0.487$ and $0.266$.
With the values of $\phi$ determined in the simulation, the PY scheme is free of adjustable parameters.
The PY-scheme solutions for $d = d_l = 1.67659$ agree best with the simulation data, which is most clearly seen 
for the higher density liquid (\figurename{}~\ref{fig:RDFs}, upper panel). 
Note that the chemical distance range in \figurename{}~\ref{fig:RDFs} is the same as
in \figurename{}~\ref{fig:Exponent}, where the fractal-dimensional cluster scaling is validated.

We have also simulated 2D hard disks with center-of-mass positions
confined to a fractal configuration space of dimension $d<2$ (results not shown).
This constitutes an alternative fractal liquid model system, where particle dimension differs from
configuration space dimension. Pair correlations in such liquids differ distinctly from the
predictions of the fractal-dimensional PY scheme. This can be rationalized by noting that
the PY-scheme equations are based on one and only one measure of distance, which is incapable of 
describing both a fractal-space correlation and an embedding-space interaction.  

In conclusion, we have simulated a fractal liquid in thermodynamic equilibrium
and demonstrated that the measured particle correlations are well predicted by
a fractal liquid integral equation.
Our approach may be easily applied to other configuration spaces with different fractal dimension.
Particle interactions beyond the simple no overlap constraint should soon be studied,
including short-ranged attraction which should result in liquid-vapor demixing, and also long-ranged repulsion.
We have reported here on a monodisperse model system, but interaction-polydispersity is already implemented in
the liquid integral equation solver for arbitrary dimension.

Fractal liquids lead the way for a plethora of new fundamental research.
At present it is unclear which types of thermodynamic phases exist in fractal dimensions and phase diagrams await to be outlined.
Field theories like density functional theory could be generalized to non-integer dimension.
Mode coupling theory for the kinetic glass transition of three-dimensional
liquids in porous media \cite{Krakoviack2005} and in arbitrary dimensions \cite{Ikeda2010, Charbonneau2011} relies on static
structure input, which can be calculated using fractal liquid integral equations.
Future studies should cover time-resolved dynamics of fractal liquids in and out of equilibrium.
Transport coefficients of fractal liquids may be studied, requiring an account for fractal hydrodynamics.
A promising application for fractal liquid theory is the prediction of thermodynamic properties of
microphase separated liquids in porous media as encountered in natural oil and gas reservoirs.

\begin{acknowledgements}
We thank Matilde Marcolli, J\"{u}rgen Horbach, Stefan U. Egelhaaf, Charles G. Slominski, Ahmad K. Omar and Mu Wang  
for numerous discussions that helped to develop the ideas presented here.
This work was supported by the ERC Advanced Grant INTERCOCOS (Grant No. 267499) and by the graduate school POROSYS.
M.H. acknowledges support by a fellowship within the Postdoc-Program of the German Academic Exchange Service 
(DAAD).
\end{acknowledgements}

\end{document}